\documentclass[prb,showpacs,twocolumn,preprintnumbers,
amsmath,groupedaddress,amssymb,superscriptaddress,floatfix]{revtex4}
\usepackage[pdftex]{graphicx}
\usepackage[latin1]{inputenc}
\usepackage{xcolor,bm}
\usepackage{nicefrac}

\begin{document}
\draft

\hyphenation{a-long}

\title{Common effect of chemical and external pressures
on the magnetic properties of RECoPO (RE = La, Pr)}

\author{G.~Prando}\email[E-mail: ]{g.prando@ifw-dresden.de}
\affiliation{Leibniz-Institut f\"ur Festk\"orper- und
Werkstoffforschung (IFW) Dresden, D-01171 Dresden, Germany}
\author{P.~Bonf\`a}
\affiliation{Dipartimento di Fisica and Unit\`a CNISM di Parma,
Universit\`a di Parma, I-43124 Parma, Italy}
\author{G.~Profeta}
\affiliation{Department of Physical and Chemical Sciences and
SPIN-CNR, Universit\`a dell'Aquila, I-67100 L'Aquila, Italy}
\author{R.~Khasanov}
\affiliation{Laboratory for Muon Spin Spectroscopy, Paul Scherrer
Institut, CH-5232 Villigen PSI, Switzerland}
\author{F.~Bernardini}
\affiliation{IOM-CNR and Dipartimento di Fisica, Universit\`a di
Cagliari, I-09042 Monserrato (Ca), Italy}
\author{M.~Mazzani}
\affiliation{Dipartimento di Fisica and Unit\`a CNISM di Parma,
Universit\`a di Parma, I-43124 Parma, Italy}
\author{E. M.~Br\"uning}
\affiliation{Leibniz-Institut f\"ur Festk\"orper- und
Werkstoffforschung (IFW) Dresden, D-01171 Dresden, Germany}
\author{A.~Pal}
\affiliation{National Physical Laboratory (CSIR), New Delhi 110012,
India}
\author{V. P. S.~Awana}
\affiliation{National Physical Laboratory (CSIR), New Delhi 110012,
India}
\author{H.-J.~Grafe}
\affiliation{Leibniz-Institut f\"ur Festk\"orper- und
Werkstoffforschung (IFW) Dresden, D-01171 Dresden, Germany}
\author{B.~B\"uchner}
\affiliation{Leibniz-Institut f\"ur Festk\"orper- und
Werkstoffforschung (IFW) Dresden, D-01171 Dresden,
Germany}\affiliation{Institut f\"ur Festk\"orperphysik, Technische
Universit\"at Dresden, D-01062 Dresden, Germany}
\author{R.~De Renzi}
\affiliation{Dipartimento di Fisica and Unit\`a CNISM di Parma,
Universit\`a di Parma, I-43124 Parma, Italy}
\author{P.~Carretta}
\affiliation{Dipartimento di Fisica and Unit\`a CNISM di Pavia,
Universit\`a di Pavia, I-27100 Pavia, Italy}
\author{S.~Sanna}
\affiliation{Dipartimento di Fisica and Unit\`a CNISM di Pavia,
Universit\`a di Pavia, I-27100 Pavia, Italy}

\widetext

\begin{abstract}
We report a detailed investigation of RECoPO (RE = La, Pr) and
LaCoAsO materials performed by means of muon spin spectroscopy.
Zero-field measurements show that the electrons localized on the
Pr$^{3+}$ ions do not play any role in the static magnetic
properties of the compounds. Magnetism at the local level is indeed
fully dominated by the weakly-itinerant ferromagnetism from the Co
sublattice only. The increase of the chemical pressure triggered by
the different ionic radii of La$^{3+}$ and Pr$^{3+}$, on the other
hand, plays a crucial role in enhancing the value of the magnetic
critical temperature and can be mimicked by the application of
external hydrostatic pressure up to $24$ kbar. A sharp discontinuity
in the local magnetic field at the muon site in LaCoPO at around $5$
kbar suggests a sizeable modification in the band structure of the
material upon increasing pressure. This scenario is qualitatively
supported by \emph{ab-initio} density-functional theory
calculations.
\end{abstract}


\date{\today}

\maketitle

\narrowtext

\section{Introduction}

The discovery of high-$T_{\textrm{c}}$ superconductivity in F-doped
REFeAsO ($1111$) was one of the most significant breakthroughs in
the field of condensed matter physics during the last
decades.\cite{Kam08,Joh10} A striking worldwide interest has been
indeed recently devoted to the synthesis of this class of materials,
looking for suitable chemical substitutions of rare-earth (RE) ions,
transition metals (TM) and pnictogen elements (Pn) allowing to
enhance the critical temperature $T_{\textrm{c}}$, the highest value
$T_{\textrm{c}} \simeq 55$ K being currently reached in
SmFeAsO$_{1-x}$F$_{x}$.\cite{Ren08,Joh10,Pra11} Electron doping can
be realized directly on the FeAs layers also leading in turn to
superconductivity. In this respect, one of the most studied chemical
substitution of TM is Fe$_{1-x}$Co$_{x}$. This applies to Fe-based
materials belonging both to $1111$ and to $122$ families, the parent
compound for the latter case being
BaFe$_{2}$As$_{2}$.\cite{Sef08a,Mar10,Awa10a,Sha12a,Sha12b,Sef08b,Yam09,Pra12a}

Beyond superconductivity, $1111$ materials show interesting magnetic
features associated with the mutual interaction among localized
electrons onto the external shells of RE ions and itinerant carriers
from the TM sublattice.\cite{Mae12} A strong $f-d$ hybridization was
shown to be present, for instance, in superconducting
SmFeAsO$_{1-x}$F$_{x}$ under conditions of optimal
doping.\cite{Pra10} In the case of undoped REFeAsO compounds, a spin
density wave (SDW) phase below $T_{\textrm{N}} \simeq 140$ K
associated with itinerant $d$ electrons from Fe is found to coexist
with an antiferromagnetically (AFM) ordered phase of RE magnetic
moments at much lower temperatures ($T_{\textrm{N}}^{\textrm{RE}}
\sim 5 - 10$ K).\cite{Mae09,DeR12} It should be stressed how the
full substitution of Fe by Co has attracted particular interest e.
g. in RECoAsO compounds where much more complex magnetic behaviours
were shown by means of both macroscopic and local experimental
techniques.\cite{Yan08,Oht09a,Oht09b,Oht10a,Oht10b,Sar10,Awa10b,Pal11a,Oht11,Sug11}
The Co sublattice, in particular, is known to enter a
weakly-itinerant ferromagnetic (FM) phase below $T_{\textrm{C}}
\simeq 60 - 80$ K, the precise value of $T_{\textrm{C}}$ being
strongly dependent on RE.\cite{Yan08,Oht09b,Oht10a,Oht11,Sug11} At
lower temperatures, the occurrence of other phase transitions is
clearly observed in samples containing magnetic RE ions as a
consequence of the strong interaction among the two magnetic
sublattices. This behaviour, in fact, is typically interpreted as
the result of progressive FM-AFM transitions of the Co sublattice
induced by the RE ions followed at the lowest temperatures by the
full magnetic ordering of the RE
sublattice.\cite{Oht09b,Oht10a,Oht11,Sug11}

A similar phenomenology has been recently reported in the
isostructural P-based RECoPO compounds.\cite{Yan08,Kre09,Pal11b}
LaCoPO has been investigated by means of {}$^{31}$P nuclear magnetic
resonance (NMR) showing that Moriya's theory of self-consistently
renormalized spin fluctuations for weakly-itinerant magnets
($T_{\textrm{C}} \simeq 35$ K) well describes the experimental
results.\cite{Maj09,Sug09,Maj10} Magneto-transport, dc magnetometry
and NMR measurements on compounds with magnetic RE ions like
Sm$^{3+}$ and Nd$^{3+}$ show the occurrence of multiple magnetic
phase transitions similarly to the case of
RECoAsO.\cite{Pal11b,Maj12} In RECoPO compounds an interaction
between the two magnetic sublattices much stronger than what is
occurring in RECoAsO should be expected. The P/As isovalent
substitution for the Pn element, in fact, is known to introduce a
strong chemical pressure making the RE ions much closer to the
itinerant layers. In the case of CeFeAs$_{1-x}$P$_{x}$O, for
instance, these effects are known to gradually suppress the SDW
phase associated with Fe upon increasing $x$, driving at the same
time the RE sublattice from an AFM ordered phase through a FM ground
state (GS) and finally towards a Kondo-screened phase where
heavy-fermion phenomenology is
recovered.\cite{Bru08,Luo10,Sar12,Jes12}

In order to evidence how crucial the role of the chemical pressure
on the magnetic properties of RECoPO materials is, we performed
measurements of muon spin spectroscopy on these compounds. In this
paper we report on the results obtained in LaCoPO and PrCoPO, where
the local magnetism was investigated upon the application of
external hydrostatic pressure up to $24$ kbar. Remarkably, in spite
of the high value of the magnetic moment expected for the free
Pr$^{3+}$ ion ($\mu_{\textrm{Pr}} \simeq 3.6 \; \mu_{\textrm{B}}$),
electrons localized on the external shells of Pr$^{3+}$ do not play
any role in the local static magnetic properties of PrCoPO probed by
muons. For both compounds, only the itinerant ferromagnetism from
the Co bands dominates the observed response. The chemical pressure
triggered by the full Pr$^{3+}$/La$^{3+}$ substitution, on the other
hand, has a sizeable effect in enhancing the value of the critical
temperature of the itinerant ferromagnetic phase. In this respect,
external hydrostatic pressure is shown to lead to the same result as
Pr/La substitution both in LaCoPO and, to a lesser extent, in the
isostructural compound LaCoAsO. Furthermore, both chemical and
external pressures strongly suppress the local magnetic field at the
muon site while leaving the magnetic moment per Co ion substantially
unchanged. It should be remarked that the chemical shrinkage of the
lattice is intrinsically expected to be characterized by a higher
degree of non-hydrostaticity that, in the case of $122$ systems, was
shown to play a drastic role in governing the resulting magnetic
properties. \cite{Col11} However, claims of close analogies among
the effect of chemical and external pressures were reported
concerning both $1111$ and $122$ compounds.\cite{Kim09,Pra12b,Gat12}
The results presented in this paper further confirm this latter
scenario also for Co-based $1111$ materials. \emph{Ab initio}
density-functional theory calculations have been developed to
describe the effects of pressure both on the interstitial
crystallographic sites for the muons and on the electronic bands of
LaCoPO. Results support the experimental findings and suggests that
chemical and external pressures can both trigger a change in the
electronic band structure. As a result, chemical and external
hydrostatic pressures act similarly on the magnetic properties of
LaCoPO.

\section{Experimental details}\label{SectChem}

\begin{figure}[t!]
\begin{center}
\includegraphics[scale=0.32]{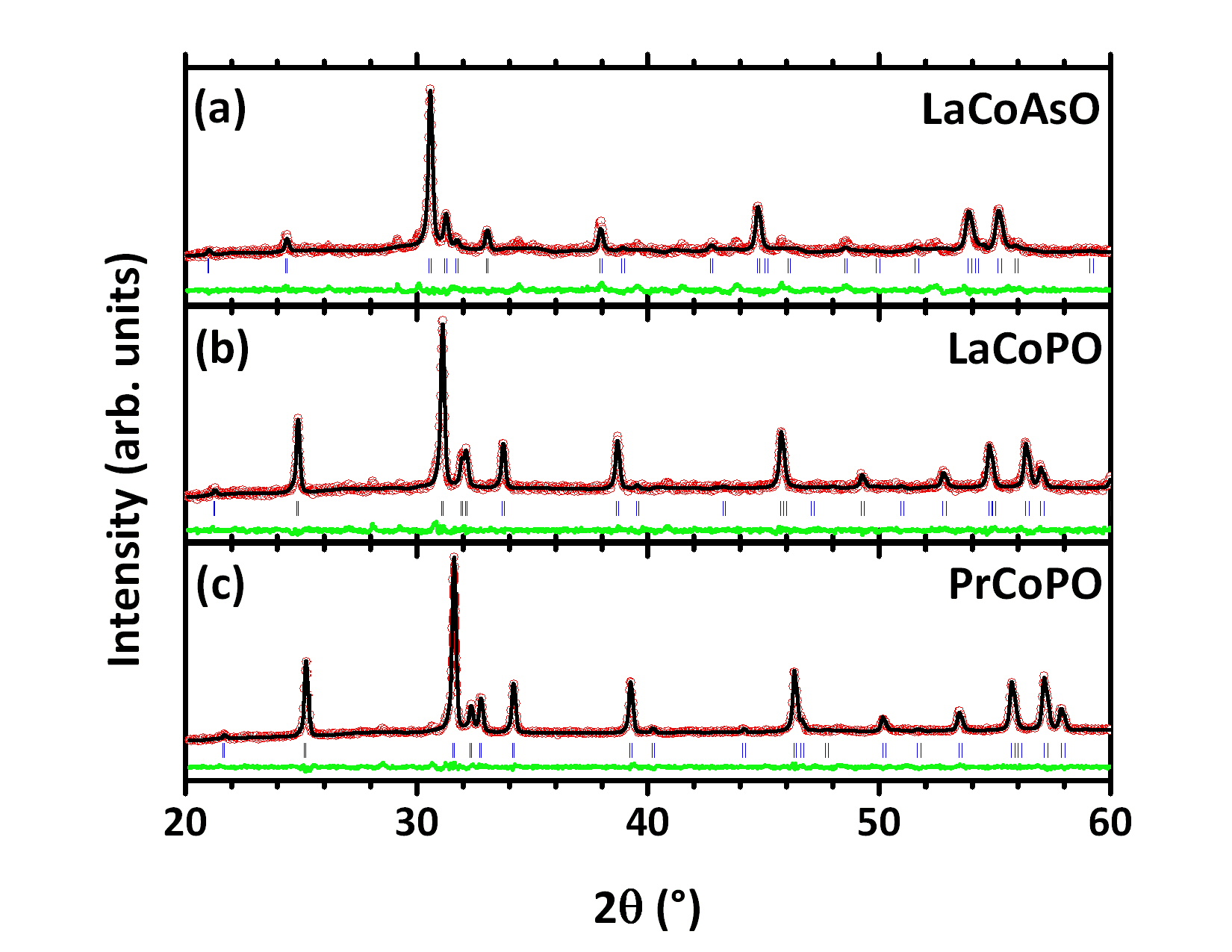}%
\caption{\label{GraXRDpatterns}(Color online) Observed (red circles)
and calculated (blue solid lines) X-ray powder-diffraction patterns
at room temperature for the investigated samples of LaCoAsO (see (a)
panel), LaCoPO (see (b) panel) and PrCoPO (see (c) panel). Black
lines are best-fits to experimental data according to a Rietveld
analysis.}
\end{center}
\end{figure}

Loose powders of LaCoAsO, LaCoPO and PrCoPO were grown via
solid-state reactions as described in detail in Refs.
\onlinecite{Pal11a} and \onlinecite{Pal11b}. The structural
properties of the lattices were measured at room temperature by
means of a Rigaku X-ray diffractometer with Cu K$_{\alpha}$
radiation. The Rietveld analysis of the diffraction patterns (see
Fig. \ref{GraXRDpatterns}) confirmed that all the samples
crystallized in the tetragonal phase, space group $P4/nmm$, and
allowed us to extract the values of the lattice parameters $a$ and
$c$ (see Tab. \ref{TabLatticeConstants}). It is clear from Tab.
\ref{TabLatticeConstants} how both the P/As and Pr/La substitutions
increase the chemical shrinkage of the cell and reduce both $a$ and
$c$ accordingly. This is in agreement with what is generally
reported for all the $1111$ family of
compounds.\cite{Oht09b,Luo10,Nit10}

\begin{table}[b!]
\caption{Lattice parameters for the investigated samples after
Rietveld refinements of X-ray powder diffraction patterns displayed
in Fig. \ref{GraXRDpatterns}.}
\label{TabLatticeConstants}%
\bgroup
    \begin{tabular}{|c|c|c|c|c|}
        \hline
        \textbf{Compound} & \textbf{$a$ (Å)} &
        \textbf{$c$ (Å)} & \textbf{Volume (Å$^{3}$)} &
        $\chi^{2}$\\
        \hline
        \hline
        LaCoAsO & $4.048(8)$ & $8.462(7)$ & $138.73(1)$ &
        $3.32$\\
        \hline
        LaCoPO & $3.966(7)$ & $8.365(0)$ & $131.62(7)$ &
        $2.12$\\
        \hline
        PrCoPO & $3.9208(4)$ & $8.212(4)$ & $126.25(0)$ &
        $2.31$\\
        \hline
    \end{tabular}
\egroup
\end{table}
Measurements of zero-magnetic-field (ZF) muon spin spectroscopy
($\mu^{+}$SR) were performed at the GPD spectrometer ($\mu$E1
beamline) of the S$\mu$S muon source at the Paul Scherrer Institut,
Switzerland (see Ref. \onlinecite{Yao11} for a comprehensive
introduction to $\mu^{+}$SR techniques). Pressure ($P$) was applied
at ambient temperature ($T$) in a double-wall piston-cylinder cell
made of MP$35$N alloy and its value was quantified by ac
susceptometry at $T \sim 3$ K from the shift of the superconducting
critical temperature of a small In wire inside the cell. The
transmitting medium Daphne oil 7373 was employed assuring that the
pressure conditions were always nearly hydrostatic in the
experimental range.\cite{Kha11,Dun10} The maximum $P$ value
attainable at low $T$ with this setup is close to $24$ kbar. The
pressure cell (PC) intrinsically leads to an extremely high
background level in ZF-$\mu^{+}$SR measurements, whose behaviour as
a function of $T$ was tested and measured in an independent set of
experiments. For this reason, a ZF-$\mu^{+}$SR characterization of
the samples was also performed at the low-background spectrometers
Dolly and GPS ($\pi$E1 and $\pi$M3 beamlines, respectively) at PSI.

dc magnetization measurements were performed by using the commercial
superconducting quantum interference device (SQUID) magnetometer
MPMS-XL7 (Quantum Design). A piston-cylinder CuBe PC (EasyLab Mcell
10) was used to apply hydrostatic $P \leq 11$ kbar. Again, the
Daphne oil 7373 was employed as transmitting medium. $P$ was applied
at ambient $T$ and its value quantified at low $T$ from the shift of
the superconducting critical temperature of a small Sn wire inside
the cell.

\section{Experimental results}\label{SectResults}

The measured spin depolarization functions for the implanted muons
($\mu^{+}$) as a function of time $\left(t\right)$ in ZF conditions
were fitted for all the samples and at all the investigated $T$
values by the general expression
\begin{equation}\label{EqGeneralFittingZFPCandSample}
    A_{T}(t) = A_{0} \left[a_{\textrm{PC}} \;
    e^{-\frac{\sigma_{\textrm{PC}}^{2} t^{2}}{2}} +
    \left(1 - a_{\textrm{PC}}\right)
    G_{T}^{\textrm{s}}(t)\right].
\end{equation}
Here the amplitude $a_{\textrm{PC}}$ accounts for the fraction of
incoming $\mu^{+}$ stopping in the PC. The depolarization of these
$\mu^{+}$ is a Gaussian-like relaxation governed by a nearly
$T$-independent $\sigma_{\textrm{PC}}$ term arising from nuclear
magnetism inside the MP$35$N alloy. The remaining fraction $\left(1
- a_{\textrm{PC}}\right)$ of $\mu^{+}$ is implanted directly into
the sample and, accordingly, the relative depolarization
$G_{T}^{\textrm{s}}(t)$ can be described as
\begin{eqnarray}\label{EqGeneralFittingZFSample}
    G_{T}^{\textrm{s}}(t) & = &
    \left[1 - V_{\textrm{m}}(T)\right]
    e^{-\frac{\sigma_{\textrm{N}}^{2}
    t^{2}}{2}} +{}\nonumber\\ & + & \left[a^{\perp}(T) F(t)
    D^{\perp}(t) + a^{\parallel}(T) D^{\parallel}(t)\right].
\end{eqnarray}
Here the quantity $V_{\textrm{m}}(T)$ represents the fraction of
$\mu^{+}$ experiencing a static local magnetic field or,
equivalently, the magnetic volume fraction of the investigated
sample. In the paramagnetic limit, namely $V_{\textrm{m}}(T) = 0$,
no static field of electronic origin contributes to the
depolarization and only the weak contribution from the nuclear
magnetic moments leads to a slow Gaussian depolarization with
characteristic rate $\sigma_{\textrm{N}}$ (typical measured values
$\sigma_{\textrm{N}} \sim 0.1 \; \mu$s$^{-1}$). Below the
magnetic-order critical transition temperature $T_{\textrm{C}}$, the
superscript $\perp$ ($\parallel$) refers to $\mu^{+}$ experiencing a
local static magnetic field in a perpendicular (parallel) direction
with respect to the initial $\mu^{+}$ spin polarization. The
amplitudes $a^{\perp,\parallel}$ must then satisfy the condition
$\left[a^{\perp}(T) + a^{\parallel}(T)\right] = V_{\textrm{m}}(T)$
accordingly. In the presence of a long-range magnetic order inside
the sample, a coherent precession of $\mu^{+}$ around the local
magnetic field $B_{\mu}$ can be discerned in the $a^{\perp}$
amplitude and described by the oscillating function $F(t)$. The
(either exponential or Gaussian) damping function $D^{\perp}(t)$
reflects a distribution of local magnetic field values at the
$\mu^{+}$ site. The $a^{\parallel}$ component, on the other hand, is
typically damped by the exponentially-decaying function
$D^{\parallel}(t) = e^{-\lambda^{\parallel}t}$ probing
spin-lattice-like relaxation processes ($\lambda^{\parallel} \sim
0.1 \; \mu$s$^{-1}$).

Standard oscillating functions $F(t) = \cos\left(\gamma B_{\mu} t +
\phi\right)$, where $\gamma = 2 \pi \times 135.54$ MHz/T is the
magnetogyric ratio for $\mu^{+}$, are in good agreement with
experimental data (statistical $\chi^{2} \simeq 1 - 1.2$). Negative
values for the phase $\left|\phi\right| \sim 20° - 30°$ were
systematically measured at the low-background spectrometer Dolly,
similarly to what reported in literature for RECoAsO
compounds.\cite{Sug11} Phase values close to $-30°$ are generally
accepted to be an experimental evidence of magnetically-ordered
phases incommensurate with the underlying crystalline lattice.

The three main contributions to the local field at the muon site
$B_{\mu}$ in the case of ferromagnetic materials come from the
dipolar field, the transferred hyperfine field and the Lorentz
field.\cite{Tra11} As it will be discussed in more detail later in
Sect. \ref{SectFPCalc} (see Eq. \eqref{EqSumLocalFields}), the
crucial physical parameters governing the amount of those
contributions are the distance between the $\mu^{+}$ and the
magnetic ions, the magnetic moment of the ordered magnetic phase and
the density of spins at the $\mu^{+}$ site.\cite{Bar86} Once the
interstitial crystallographic position of the $\mu^{+}$ is known,
the independent knowledge of the value of the magnetic moment (by
means of e. g. dc magnetometry) is of crucial importance since it
allows one to directly access the transferred hyperfine
field.\cite{Tra11,Bar86}

\begin{figure}[t!]
\begin{center}
\includegraphics[scale=0.32]{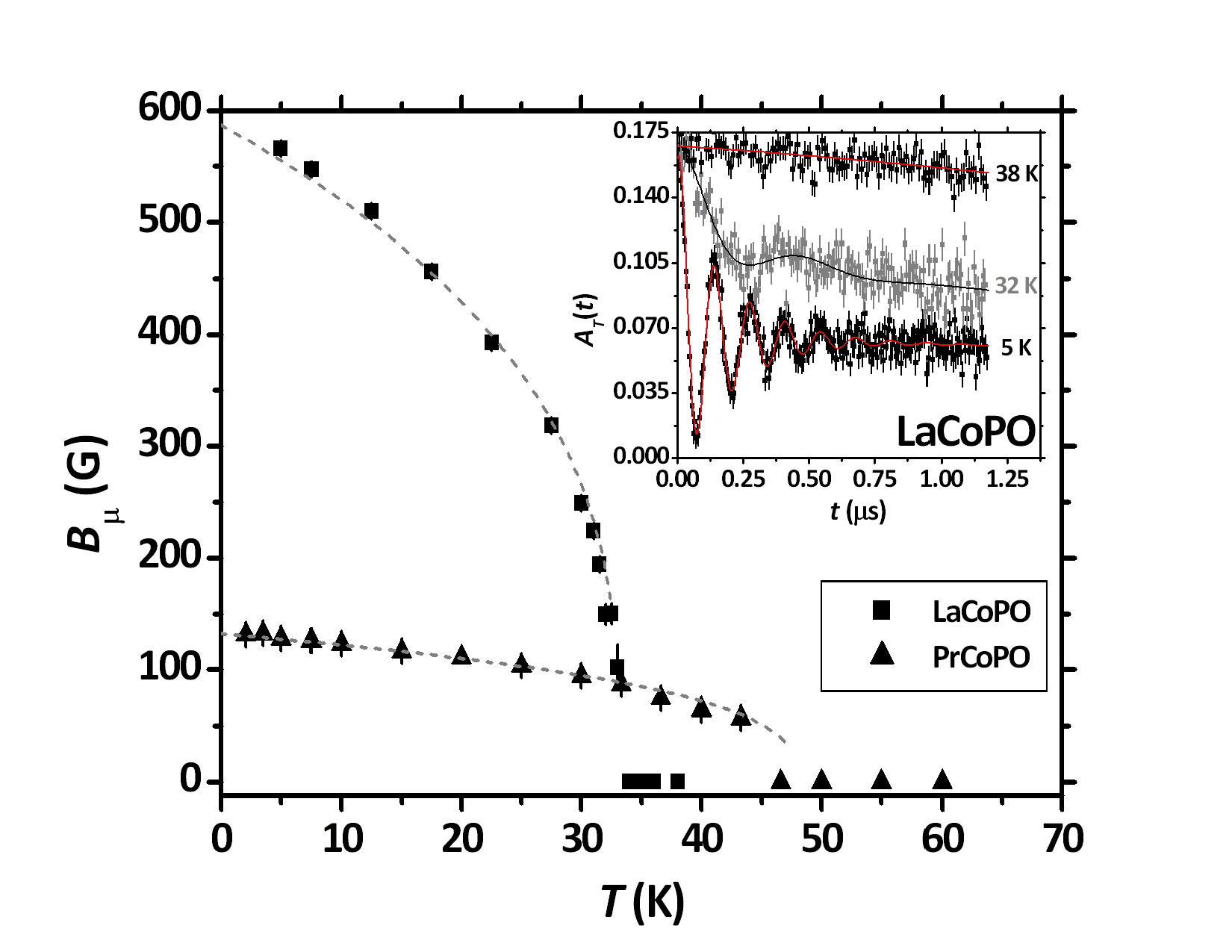}%
\caption{\label{GraInternalFields}(Color online) Main experimental
results of ZF-$\mu^{+}$SR (ambient $P$, Dolly spectrometer). Main
panel: $B_{\mu}$ vs. $T$ for LaCoPO and PrCoPO. Dashed lines are
best-fits to experimental data according to Eq.
\ref{EqInternalFieldMeanField}. Inset: raw ZF spin depolarization
data for LaCoPO at some selected $T$ values. Continuous lines are
best fits to experimental data according to Eq.
\ref{EqGeneralFittingZFPCandSample}, where $a_{\textrm{PC}} = 0$.}
\end{center}
\end{figure}

Representative ZF $\mu^{+}$-depolarizations for LaCoPO obtained at
Dolly (ambient $P$) are presented in the inset of Fig.
\ref{GraInternalFields}. Coherent oscillations indicative of the
establishment of a long-range magnetic order can be clearly
discerned for $T \leq T_{\textrm{C}} \simeq 33$ K down to the lowest
investigated temperature. Experimental data were fitted by means of
Eq. \eqref{EqGeneralFittingZFPCandSample} accordingly, where
$a_{\textrm{PC}} = 0$. The $T$-dependence of the magnetic volume
fraction $V_{\textrm{m}}$ (not shown) confirms that the sample is
fully magnetic below $T_{\textrm{C}}$. The $B_{\mu}$ vs. $T$ trend
is plotted in the main panel of Fig. \ref{GraInternalFields} and can
be fitted according to a power-law function
\begin{equation}\label{EqInternalFieldMeanField}
    B_{\mu}(T) =
    B_{\mu}(0)\left(1-\frac{T}{T_{\textrm{C}}}\right)^{\beta}
\end{equation}
eventually saturating at $B_{\mu}(0) = 587.5 \pm 2.5$ G, the
exponent turning out to be $\beta = 0.34 \pm 0.01$.

Also in the case of PrCoPO, measurements performed at Dolly reveal a
static long-range magnetism probed by $\mu^{+}$ in $100$\% of the
magnetic volume (not shown) below $T_{\textrm{C}} \simeq 48$ K. The
value for $T_{\textrm{C}}$ is sizeably increased with respect to
what is observed in LaCoPO, consistently with what is reported for
RECoAsO compounds.\cite{Sug11} Remarkably, the $B_{\mu}$ vs. $T$
trend still can be well-described by Eq.
\eqref{EqInternalFieldMeanField} where again $\beta = 0.34 \pm 0.01$
(see the main panel of Fig. \ref{GraInternalFields}). In the case of
PrCoPO, only the local field $B_{\mu}(0) = 132 \pm 2$ G is more than
$4$ times lower than in LaCoPO. In spite of the high value for the
localized magnetic moment expected for the free Pr$^{3+}$ ion ($\sim
3.6 \; \mu_{\textrm{B}}$), the results for the PrCoPO sample are
qualitatively identical to the case of LaCoPO. This is clearly
different from what was previously reported in the literature for
the As-based compound PrCoAsO, where a more complex phenomenology
for ZF-$\mu^{+}$SR (displaying a splitting of internal fields) was
observed.\cite{Sug11} Nevertheless, dc magnetization measurements
for LaCoAsO and PrCoAsO were reported to display no sizeable
differences among themselves, no anomalies being detected for $T
\leq T_{\textrm{C}}$.\cite{Sug11} The present results (together with
what is reported in Ref. \onlinecite{Sug11}) then suggest that both
in PrCoPO and in PrCoAsO the magnetic moments localized on the
Pr$^{3+}$ ions do not play any role in the static magnetic features
of these compounds. The phenomenology observed in
PrCoAsO\cite{Sug11} then can possibly be due to the presence of two
different $\mu^{+}$ sites whose statistical population is strongly
dependent on $T$ in the investigated experimental range.

The overall scenario deduced from our measurements on LaCoPO and
PrCoPO and from the previous characterization of the materials (see
Sect. \ref{SectChem}) suggests that the main effect of the full
Pr/La substitution is only associated with a lattice shrinkage
induced by the different chemical pressures associated to the
different ionic radii of La$^{3+}$ and Pr$^{3+}$. This is different
from the cases of all other magnetic RE ions where the interplay of
the localized and the itinerant magnetic degrees of freedom leads to
progressive FM-AFM re-ordering effects on the Co sublattice.
Therefore, it seems safe to deduce that Pr$^{3+}$ magnetic moments
do not contribute to the local static magnetic properties of the
compound that are only governed by weakly-itinerant FM from Co.

\begin{figure}[t!]
\begin{center}
\includegraphics[scale=0.32]{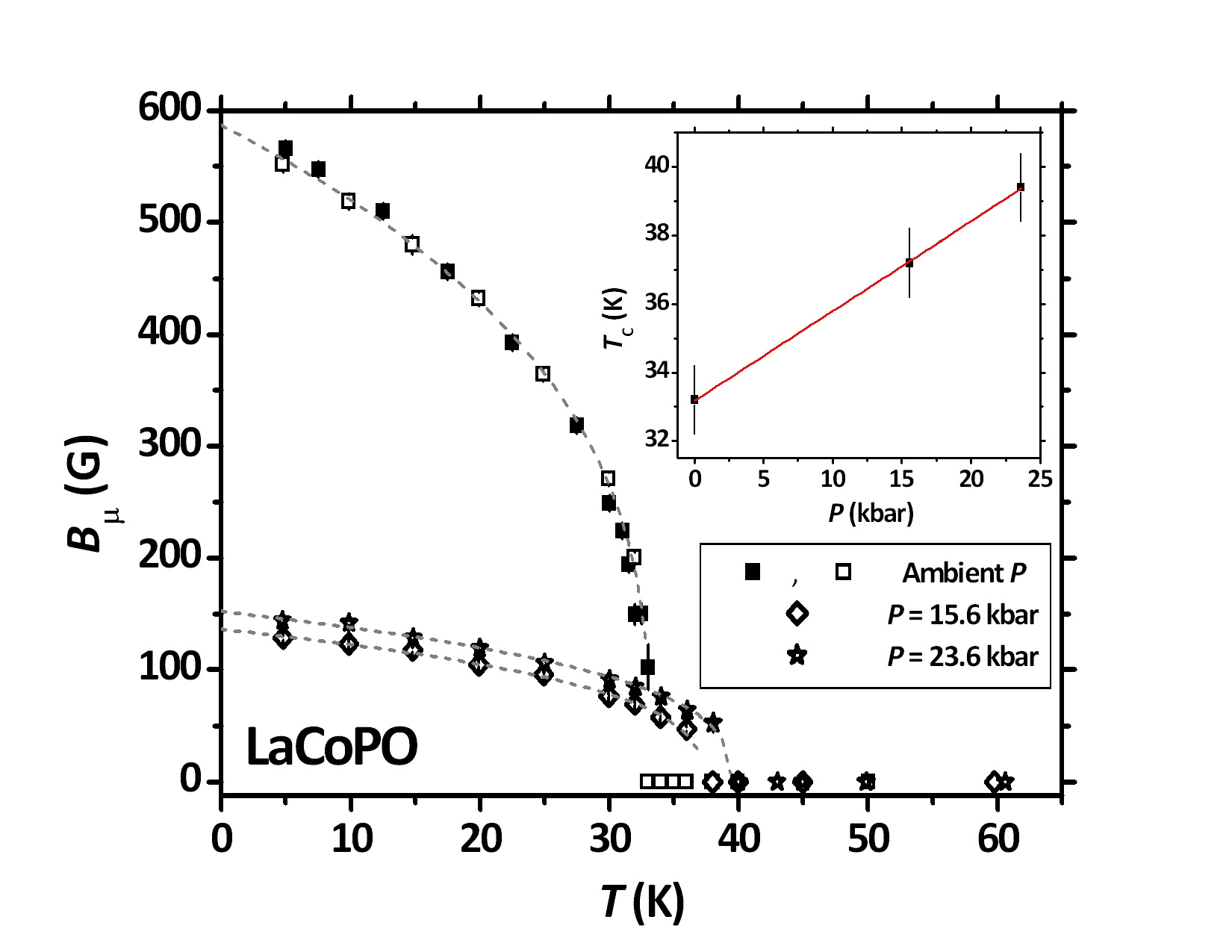}%
\caption{\label{GraInternalFieldLaCoPO} (Color online) Main panel:
$B_{\mu}(T)$ at different $P$ for LaCoPO. Dashed lines are best fits
to experimental data according to Eq. \ref{EqInternalFieldMeanField}
with $\beta = 0.34$ as fixed parameter. Close and open symbols for
ambient $P$ data refer to measurements performed at Dolly and at GPD
with unloaded cell, respectively. Inset: $T_{\textrm{C}}$ vs. $P$.
The continuous line is a best-fit to data according to a linear
function.}
\end{center}
\end{figure}

\begin{figure}[t!]
\begin{center}
\includegraphics[scale=0.32]{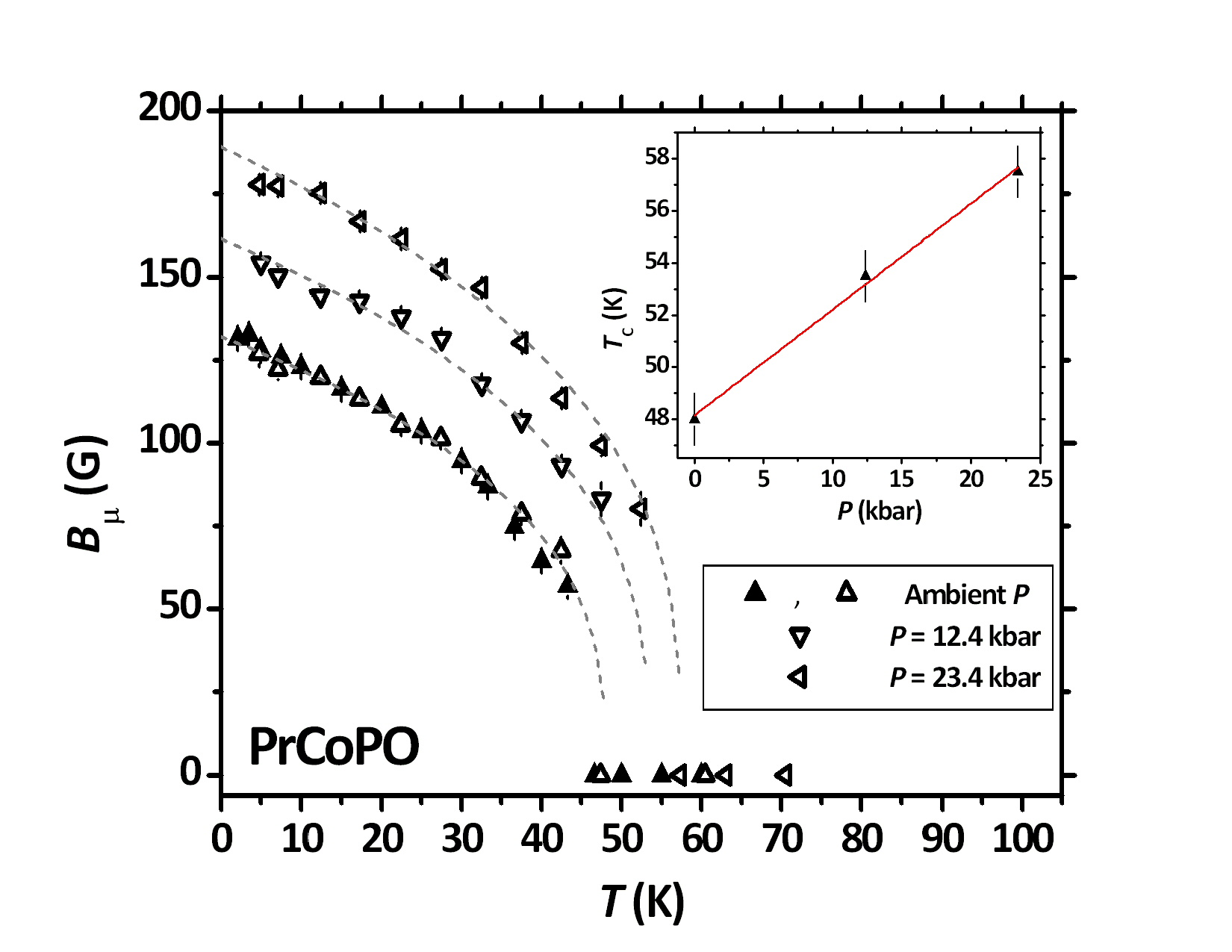}%
\caption{\label{GraInternalFieldPrCoPO}  (Color online)  Main panel:
$B_{\mu}(T)$ at different $P$ for PrCoPO. Dashed lines are best fits
to experimental data according to Eq. \ref{EqInternalFieldMeanField}
with $\beta = 0.34$ as fixed parameter. Close and open symbols for
ambient $P$ data refer to measurements performed at Dolly and at GPD
with unloaded cell, respectively. Inset: $T_{\textrm{C}}$ vs. $P$.
The continuous line is a best-fit to data according to a linear
function.}
\end{center}
\end{figure}

\begin{figure}[t!]
\begin{center}
\includegraphics[scale=0.32]{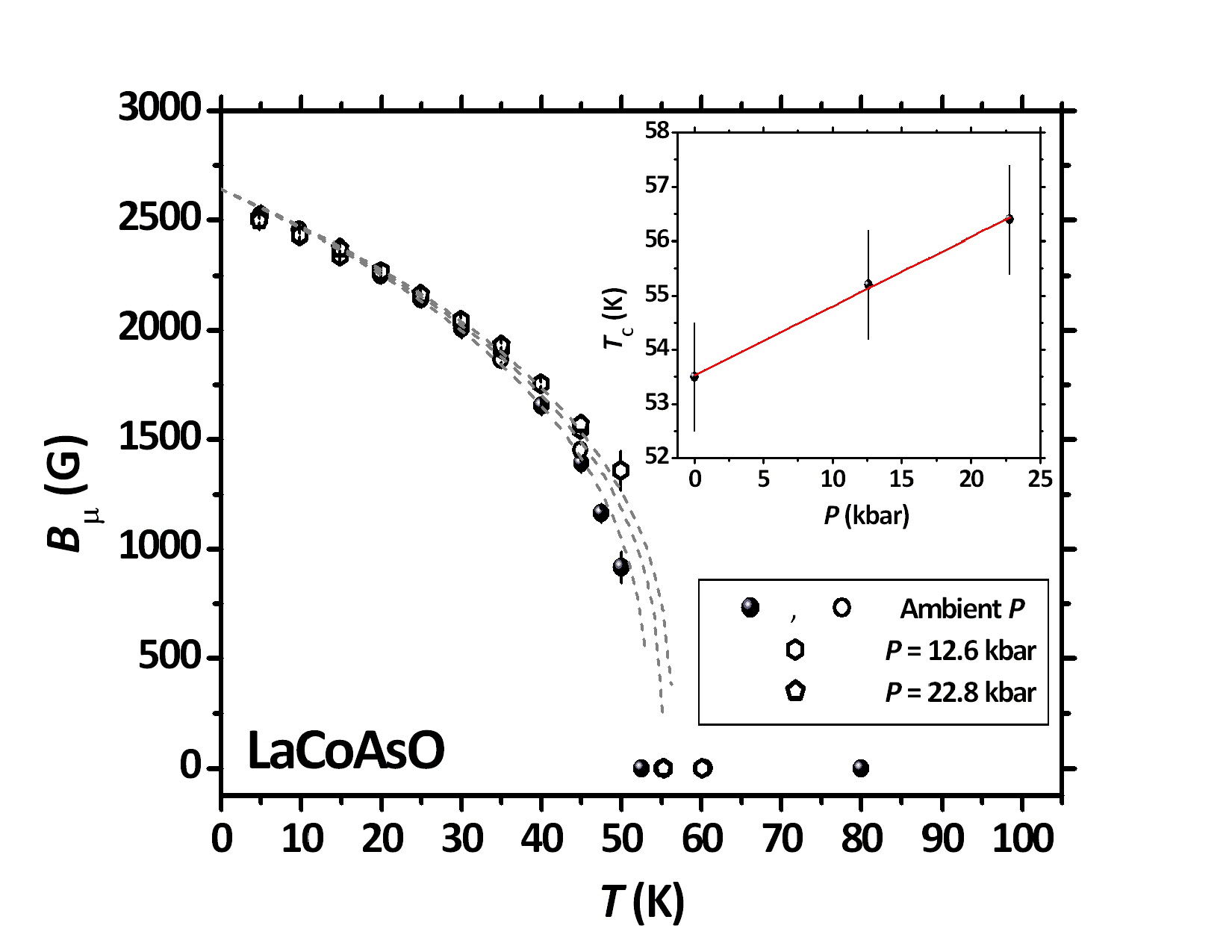}%
\caption{\label{GraInternalFieldLaCoAsO}  (Color online)  Main
panel: $B_{\mu}(T)$ at different $P$ for LaCoAsO. Dashed lines are
best fits to experimental data according to Eq.
\ref{EqInternalFieldMeanField} with $\beta = 0.34$ as fixed
parameter. Close and open symbols for ambient $P$ data refer to
measurements performed at Dolly and at GPD with unloaded cell,
respectively. Inset: $T_{\textrm{C}}$ vs. $P$. The continuous line
is a best-fit to data according to a linear function.}
\end{center}
\end{figure}

The validity of these arguments is corroborated by the measurements
under external $P$ performed at the GPD spectrometer. In spite of
the fact that almost half of the incoming $\mu^{+}$ stop inside the
pressure cell ($a_{\textrm{PC}} \sim 0.5$), clear oscillations can
be distinguished from the ZF signal associated with all the samples
below characteristic critical transition temperatures
$T_{\textrm{C}}(P)$. The analysis of $V_{\textrm{m}}$ (not shown)
clearly shows that the magnetic phase is extended over the whole
sample volume independently on the $P$ value for all the
investigated materials. The $T$-dependence of $B_{\mu}$ for the
three samples at different values of $P$ are reported in the main
panels of Figs. \ref{GraInternalFieldLaCoPO},
\ref{GraInternalFieldPrCoPO} and \ref{GraInternalFieldLaCoAsO}
(LaCoPO, PrCoPO and LaCoAsO, respectively). Here, the
$T_{\textrm{C}}$ values are estimated as free parameters from the
fitting to experimental ZF data according to Eq.
\eqref{EqInternalFieldMeanField} where, in view of the results of
ambient pressure measurements, $\beta = 0.34$ is kept as a fixed
parameter for all the samples.

By focussing on the results for LaCoPO first and by comparing the
main panels of Figs. \ref{GraInternalFields} and
\ref{GraInternalFieldLaCoPO}, it is immediate to realize that the
qualitative effect of increasing $P$ is extremely similar to what is
induced by the full Pr/La substitution. In particular, a strong
suppression of $B_{\mu}(0)$ together with a sizeable enhancement of
$T_{\textrm{C}}$ are obtained at the maximum applied $P \simeq 23.6$
kbar. As shown by the dashed lines in the main panel of Fig.
\ref{GraInternalFieldLaCoPO}, the fitting function reported in Eq.
\eqref{EqInternalFieldMeanField} still well reproduces the observed
experimental data with $\beta = 0.34$ at all the $P$ values. By more
carefully investigating the experimental results reported in Fig.
\ref{GraInternalFieldLaCoPO} one realizes that the suppression of
$B_{\mu}$ is actually non-monotonic with increasing $P$ (see also
Fig. \ref{GraIntFieldAndMagMom} later on). Remarkably, this sharp
jump in the internal magnetic field is not reflected in the
$P$-dependence of $T_{\textrm{C}}$ that steadily increases in a
linear fashion across the whole experimental range (see the inset of
Fig. \ref{GraInternalFieldLaCoPO}). Quantitative data relative to
the linear increase of $T_{\textrm{C}}$ as a function of $P$ are
reported in Tab. \ref{TabPressureSlopes}.

\begin{table}[b!]
\caption{Summarizing quantities of interest for the three
investigated compounds after ZF-$\mu^{+}$SR measurements under $P$.
$T_{\textrm{C}}(0)$ represents the critical temperature at ambient
$P$ while $\left(\nicefrac{dT_{\textrm{C}}}{dP}\right)$ is the slope
of the linear trends of $T_{\textrm{C}}$ vs. $P$ presented in the
insets of Figs. \ref{GraInternalFieldLaCoPO},
\ref{GraInternalFieldPrCoPO} and \ref{GraInternalFieldLaCoAsO}.}
\label{TabPressureSlopes}%
\bgroup
    \begin{tabular}{|c|c c c|c|}
        \hline
        \textbf{Compound} & \phantom{aa} &
        $T_{\textrm{C}}(0)$ \textbf{(K)} & \phantom{aa} &
        $\left(\nicefrac{1}{T_{\textrm{C}}(0)}\right) \cdot
        \left(\nicefrac{dT_{\textrm{C}}}{dP}\right)$
        \textbf{(kbar$^{-1}$)}\\
        \hline
        \hline
        LaCoAsO & \phantom{aa} & $53.5 \pm 1.0$ &
        \phantom{aa} &
        $\left(2.35 \pm 0.15\right) \times 10^{-3}$\\
        \hline
        LaCoPO & \phantom{aa} & $33.2 \pm 1.0$ &
        \phantom{aa} &
        $\left(8.0 \pm 0.5\right) \times 10^{-3}$\\
        \hline
        PrCoPO & \phantom{aa} & $48.0 \pm 1.0$ &
        \phantom{aa} &
        $\left(8.5 \pm 0.5\right) \times 10^{-3}$\\
        \hline
    \end{tabular}
\egroup
\end{table}

As already shown in Fig. \ref{GraInternalFields} for the
measurements at ambient $P$, the $T$-dependence of $B_{\mu}$ in
PrCoPO is qualitatively identical to what it is reported for LaCoPO.
Remarkably, measurements performed at GPD (reported in Fig.
\ref{GraInternalFieldPrCoPO}) confirm this qualitative result also
for $P$ values up to $23.4$ kbar. As it is found in the case of
LaCoPO, a linear increase of $T_{\textrm{C}}$ as a function of $P$
can be inferred (see the inset of Fig. \ref{GraInternalFieldPrCoPO}
and Tab. \ref{TabPressureSlopes}). However, differently from LaCoPO,
the saturation value of the internal field $B_{\mu}(0)$ is steadily
enhanced upon increasing $P$, reaching $B_{\mu}(0) \simeq 189$ G at
the maximum $P$ value (see also Fig. \ref{GraIntFieldAndMagMom}
later on).

A totally different phenomenology is detected for LaCoAsO up to the
value $P = 22.8$ kbar. As it is clear from Fig.
\ref{GraInternalFieldLaCoAsO}, the value $B_{\mu}(0) \simeq 2640$ G
does not display any dependence upon $P$ within the experimental
error while the linear increase of $T_{\textrm{C}}$ is much less
marked than in the case of LaCoPO and PrCoPO (see Tab.
\ref{TabPressureSlopes}). It should be remarked that, as it is
discussed in detail later in Sect. \ref{SectFPCalc}, the sizeable
difference of $B_{\mu}(0)$ between the case of P- and As-based
samples can be explained by the preferential occupation of different
crystallographic sites by the muons in the two cases (close to LaO
and Co(Pn) tri-layers, respectively). However, the function reported
in Eq. \eqref{EqInternalFieldMeanField} still yields to good fitting
results for $B_{\mu}(T)$ data with $\beta = 0.34$ as a fixed
parameter independently on the $P$ value. Since both the materials
with La as non-magnetic RE ion then share the same power-law-like
trend, it is further confirmed that this must be entirely associated
to the Co sublattice. It seems safe to deduce that this behaviour
for $B_{\mu}(T)$ is the fingerprint of static magnetism from Co in
RECo(As,P)O materials. Data at ambient $P$ are in good agreement
with what was previously reported in the literature.\cite{Sug11}

\begin{figure}[t!]
\begin{center}
\includegraphics[scale=0.32]{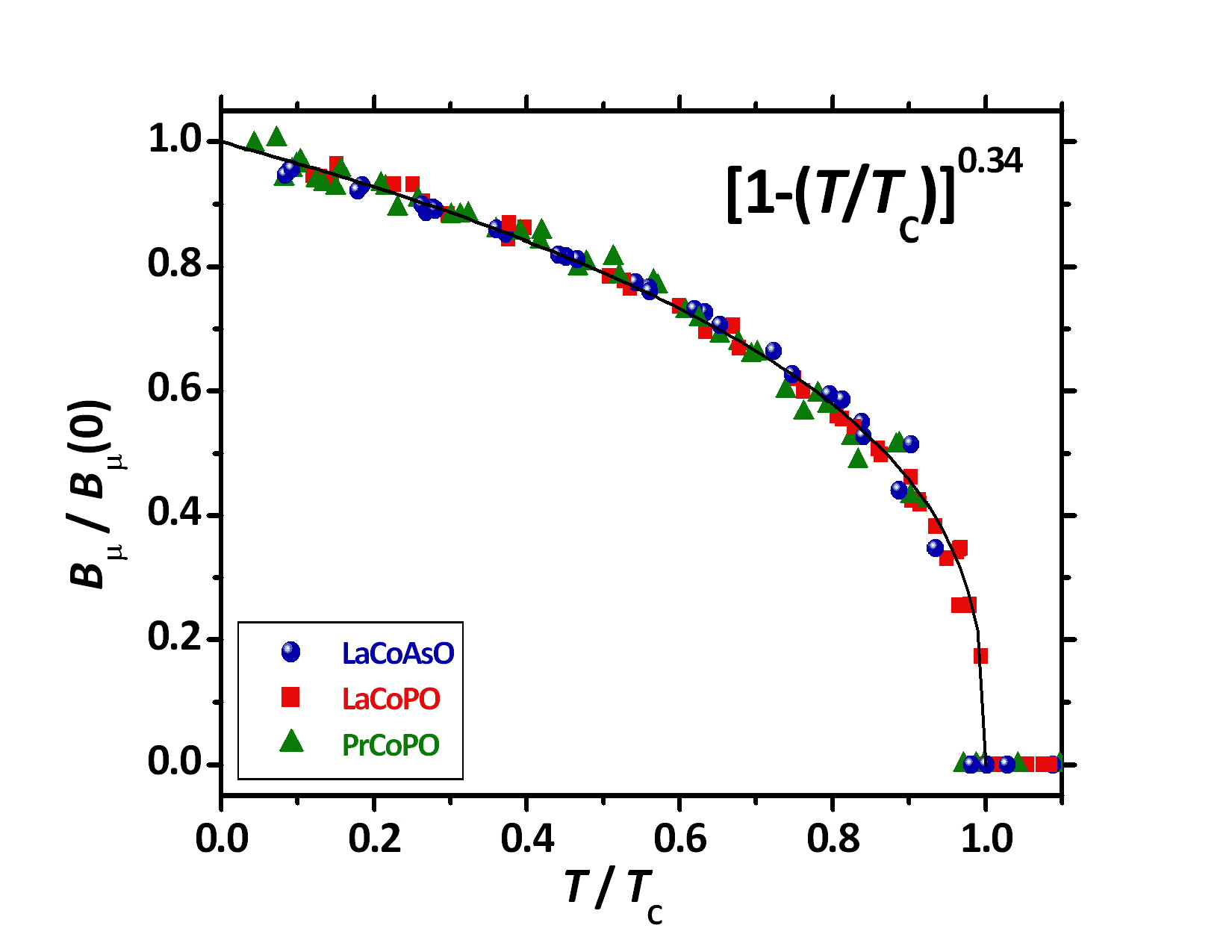}%
\caption{\label{GraCollapseOnly} (Color online) $B_{\mu}$ vs. $T$
for the three samples at all the investigated $P$ values. Data
clearly collapse on a single common power-law-like trend with $\beta
= 0.34$ (continuous line) after a proper scaling of both $T$ and
$B_{\mu}$ axes with $T_{\textrm{C}}$ and $B_{\mu}(0)$ values,
respectively.}
\end{center}
\end{figure}
\begin{figure}[b!]
\begin{center}
\includegraphics[scale=0.32]{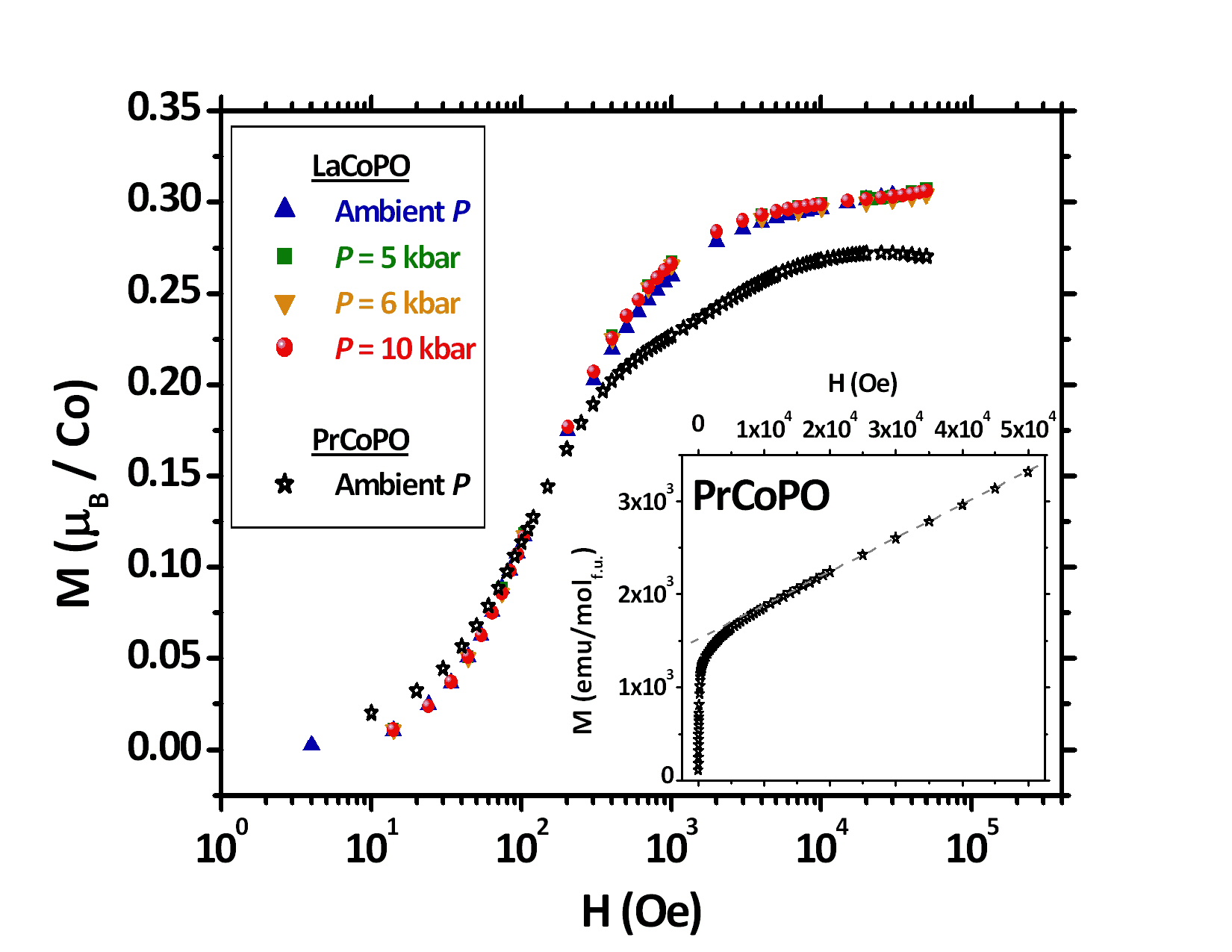}%
\caption{\label{GraSQUID} (Color online) Main panel: $M$ vs. $H$ at
fixed $T = 5$ K for LaCoPO at different $P$ values and for PrCoPO at
ambient pressure after the subtraction of the linear paramagnetic
term due to Pr$^{3+}$ ions. Inset: raw magnetization data for PrCoPO
before the subtraction of the linear paramagnetic term evidenced by
the dashed line.}
\end{center}
\end{figure}
In order to stress the strong qualitative analogies among the three
investigated samples, the experimental data displaying $B_{\mu}$ vs.
$T$ at all the different $P$ values are plotted in Fig.
\ref{GraCollapseOnly}. Here, a scaling procedure along both the $T$
and $B_{\mu}$ axes with, respectively, $T_{\textrm{C}}$ and
$B_{\mu}(0)$ values proper of each data set allows us to clearly
enlighten a shared power-law-like trend with exponent $\beta = 0.34$
common to all the samples and independent on $P$.

As already commented above (and as further discussed in detail later
in Sect. \ref{SectFPCalc}), the relative variations of $B_{\mu}$ in
ferromagnets cannot be ascribed only to variations in the value of
the magnetic moment in the ordered phase and a further experimental
study for such quantity should be independently performed. To this
aim, the behaviour of $\mu_{\textrm{Co}}$ vs. $P$ in LaCoPO was
investigated in the low-$P$ region by means of dc magnetometry.
Measurements of dc magnetization ($M$) as a function of the magnetic
field ($H$) were performed at $T = 5$ K up to $P = 10$ kbar. The
results are reported in the main panel of Fig. \ref{GraSQUID}. The
unique intrinsic contribution to $M$ in LaCoPO is expected to come
from the weakly-itinerant FM state associated with the Co sublattice
and an ordered value of $\mu_{\textrm{Co}}^{\textrm{LaCoPO}} = 0.3
\pm 0.02 \; \mu_{\textrm{B}}$ can be indeed estimated, in good
agreement with previous reports.\cite{Pal11b,Yan08} Remarkably,
$\mu_{\textrm{Co}}^{\textrm{LaCoPO}}$ turns out to be independent on
$P$ within the experimental error in the investigated $P$ range. The
same scenario holds for PrCoPO, as it is deduced from $M$ vs. $H$
measurements at ambient $P$ (raw data are presented in the inset of
Fig. \ref{GraSQUID}). After the subtraction of the linear term
enlightened in the inset of Fig. \ref{GraSQUID} by means of a dashed
line and accounting for the paramagnetic contribution of Pr$^{3+}$
magnetic moments, one obtains the intrinsic contribution of the Co
sublattice reported in the main panel of Fig. \ref{GraSQUID}.
Results only show a slight reduction of the magnetic moment of Co in
PrCoPO with respect to what was found for LaCoPO
($\mu_{\textrm{Co}}^{\textrm{PrCoPO}} = 0.275 \pm 0.025 \;
\mu_{\textrm{B}}$ should be compared with
$\mu_{\textrm{Co}}^{\textrm{LaCoPO}} = 0.3 \pm 0.02 \;
\mu_{\textrm{B}}$). These results clearly indicate that the dramatic
drop in $B_{\mu}(P)$ described in the main panel of Fig.
\ref{GraInternalFieldLaCoPO} cannot be ascribed to a sudden
suppression of $\mu_{\textrm{Co}}$.

\begin{figure}[t!]
\begin{center}
\includegraphics[scale=0.32]{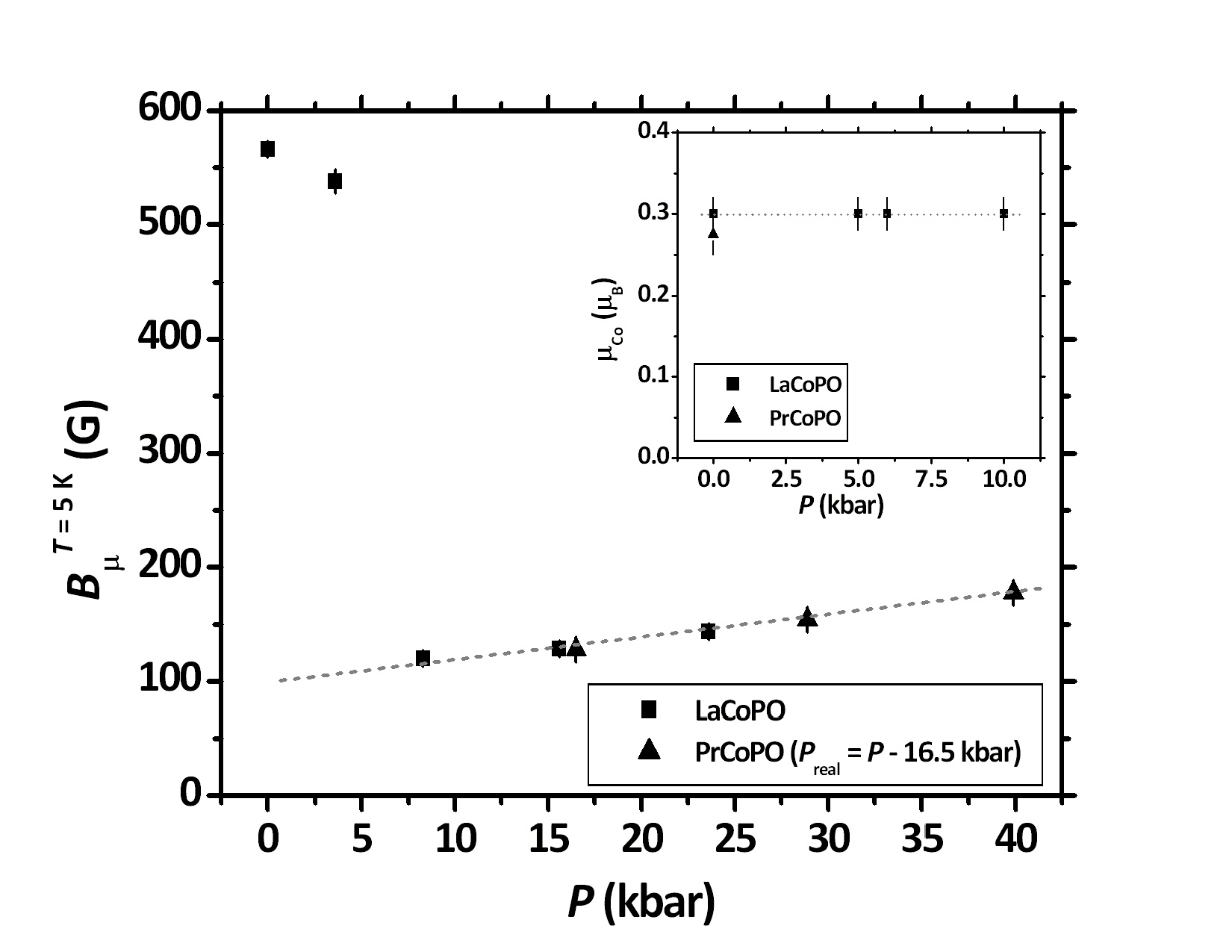}%
\caption{\label{GraIntFieldAndMagMom} (Color online) Main panel:
$B_{\mu}$ vs. $P$ at $T = 5$ K for both LaCoPO and PrCoPO. A biasing
value $P_{\textrm{B}} = 16.5$ kbar was added to real $P$ values for
PrCoPO. The dashed line is a linear guide for the eye. Inset: values
of $\mu_{\textrm{Co}}$ deduced for LaCoPO and PrCoPO from the dc
magnetization measurements presented in Fig. \ref{GraSQUID}. The
dashed line is a guide for the eye.}
\end{center}
\end{figure}
A summary of the main experimental results for $B_{\mu}$ vs. $P$ in
LaCoPO and PrCoPO at $T = 5$ K is reported in the main panel of Fig.
\ref{GraIntFieldAndMagMom}. Data already reported in Fig.
\ref{GraInternalFieldLaCoPO} for LaCoPO have been complemented by
two single measurements at $T = 5$ K and at low $P$ values ($\sim 3$
and $\sim 8$ kbar). This allows us to confirm the reproducibility of
the dramatic reduction of $B_{\mu}$ and to characterize it as a
sudden phenomenon occurring in a window with maximum width of few
kbar around $P \sim 5$ kbar. Data for PrCoPO, on the other hand,
have been plotted in the main panel of Fig.
\ref{GraIntFieldAndMagMom} after adding a phenomenological biasing
pressure $P_{\textrm{B}} = + 16.5$ kbar. It is interesting to notice
that, after such shift along the $P$ axis, a systematic common
enhancement of $B_{\mu}$ with increasing $P$ can be clearly
discerned sharing the same slope ($1.95 \pm 0.05$ G/kbar) for the
two different materials. Naively, one can think of $P_{\textrm{B}}$
as an effective pressure equivalent to the internal chemical
pressure associated with the full Pr/La substitution. The inset of
Fig. \ref{GraIntFieldAndMagMom} summarizes the behaviour of
$\mu_{\textrm{Co}}$ vs. $P$ as determined by dc magnetization
measurements. It is clear how, in the range of $P$ values up to $10$
kbar, the value of $\mu_{\textrm{Co}}$ is indeed constant and
approximately equal to $0.3$ $\mu_{\textrm{B}}$.

Modifications of contributions to the local field other than the
dipolar coupling among $\mu^{+}$ and Co layers should then be
responsible for the observed behaviour. This topic will be discussed
in more detail in Sect. \ref{SectFPCalc} from a computational point
of view. In particular, it will be shown how a sizeable effect of
$P$ in modifying the band structure and, accordingly, the
transferred hyperfine term to the $\mu^{+}$ must be taken into
consideration in order to explain the observed behaviour.

\section{Insights into the internal field at the muon site
from first-principle calculations}\label{SectFPCalc}

In this Section we will discuss the sharp suppression of
$B_{\mu}(P)$ occurring at $P = 5 \pm 2$ kbar in LaCoPO (as presented
in the main panel of Fig. \ref{GraIntFieldAndMagMom}) by taking into
consideration different possibilities explaining its origin. To this
aim, one should first write down the internal field $B_{\mu}$ at the
$\mu^{+}$ site (whose position is identified by the vector
$\mathbf{r}_{\mu}$) in a non-magnetized FM material as a sum of
several contributions. In particular, one has
\begin{equation}\label{EqSumLocalFields}
    B_{\mu} =
    \left|\textbf{B}_{\textrm{dip}}(\mathbf{r}_{\mu}) +
    \textbf{B}_{\textrm{c}}(\mathbf{r}_{\mu}) +
    \textbf{B}_{\textrm{L}}\right|
\end{equation}
where $\textbf{B}_{\textrm{dip}}(\mathbf{r}_{\mu})$ is the dipolar
field arising from the atoms within a sphere with diameter smaller
than $R_{\textrm{md}}$ ($R_{\textrm{md}}$ being the approximate
linear dimension of a magnetic domain),
$\textbf{B}_{\textrm{c}}(\mathbf{r}_{\mu})$ is the contact hyperfine
field and $\textbf{B}_{\textrm{L}}$ the Lorentz field. By
considering the value $\mu_{\textrm{Co}} \simeq 0.3 \;
\mu_{\textrm{B}}$ derived in LaCoPO (the superscript was dropped to
the aim of clarity) by means of dc magnetization measurements, it is
straightforwardly derived that $\left|\textbf{B}_{\textrm{L}}\right|
= \left(\nicefrac{4 \pi}{3}\right) M_{sat} \simeq 170$ G.

As previously discussed in Sect. \ref{SectResults}, no modification
of $\mu_{\textrm{Co}}$ could be discerned in LaCoPO upon increasing
the value of the external $P$. The drop of $B_{\mu}$ measured at $P
= 5 \pm 2$ kbar may then only be related to the quantities whose
values could still be a function of $P$, namely $\mathbf{r}_{\mu}$
and $\textbf{B}_{\textrm{c}}$. In the former case, results would
then imply that a change of the $\mu^{+}$ site is at the origin of
the observed discontinuity. Such $\mu^{+}$ jump could be possibly
due either to the lattice undergoing a structural transition or,
more generally, to a smoother modification of the electrostatic
energy landscape triggered by $P$ inducing a change in the $\mu^{+}$
preferential site. In this respect, it should be remarked that
estimates of the structural parameters as a function of $P$ by means
of X-ray diffraction measurements would be of the utmost importance
in order to get further information. In the latter case, on the
other hand, the sudden modification of the $\textbf{B}_{\textrm{c}}$
term could follow from the evolution of the structure of electronic
bands upon increasing $P$. One could also consider other
possibilities like, for instance, a $P$-induced change in the
magnetic GS from e. g. FM to AFM, or a reorientation of the magnetic
configuration in the Co sublattice upon gradually increasing $P$.
Finally, and less interestingly, another possibility is that the
experimental findings are the result of the local distortion
introduced by $\mu^{+}$ once thermalized inside the material.

In order to get more insights into the origin of the sharp drop in
$B_{\mu}$ vs. $P$ as revealed by $\mu^{+}$SR experiments on LaCoPO,
\emph{ab-initio} calculations were then carried out in order to
check for the reliability of all the scenarios proposed above for
the material under investigation. The main objects were the
computation as a function of $P$ of the crystallographic sites where
$\mu^{+}$ sit and of the energy-volume curves allowing one to derive
the structural and magnetic GSs of the system.

\subsection{Computational details}

We used both the plane wave (PW) and the full potential linearized
augmented plane waves (FP-LAPW) methods as implemented in the
VASP\cite{Kre96} and Elk\cite{Elk} packages. The
Perdew-Bruke-Ernzerhof (PBE) functional was used in order to
evaluate the exchange-correlation potential.\cite{Per96} As for the
PW approach, the electron-ion interaction was described by the
projector augmented wave (PAW) pseudopotentials method.\cite{Blo94}
Electronic convergence was set up at $10^{-6}$ eV and sampling of
the Brillouin zone performed by the Monkhorst-Pack
scheme\cite{Mon76} on a $8 \times 8 \times 4$ grid. A plane-wave
cutoff of $600$ eV and a Gaussian smearing of $0.02$ eV was used
throughout. FP-LAPW calculations were carried out using a basis set
with $R_{\textrm{min}}^{\textrm{MT}} \times \textrm{max}(\|k\|) =
7.5$, $R_{\textrm{min}}^{\textrm{MT}}$ being the smallest muffin-tin
(MT) radius inside the MT spheres, and $l_{\textrm{max}} = 8$ for
the angular momentum expansion in the MTs (for both the wave
functions and the potential). The reciprocal space was sampled with
the same grid used in the PW approach. A close agreement is observed
between the results obtained by means of both the computational
methods.

\begin{table*}[t!]
\caption{Crystallographic positions of the interstitial sites A, B
and C at two different values of $P$. The absolute values of the
local magnetic field at A, B and C arising only from the
$\mathbf{B}_{\textrm{dip}}$ contribution in Eq.
\eqref{EqSumLocalFields} are reported for two different FM
configurations of $\mu_{\textrm{Co}}$. The values of the minima of
potential energy for the three sites are denoted by $V_{0}^{i}$
while $E_{\textrm{ZP}}^{i}$ represents the lower eigenvalue for the
ZP motion of $\mu^{+}$ into the $i$ potential dip (see text). The
values for the energy barriers between different sites are denoted
as $\Delta_{i,i^{\prime}}$. All the energy values ($V_{0}^{i}$,
$E_{\textrm{ZP}}^{i}$ and $\Delta_{i,i^{\prime}}$) are
conventionally referred to $V_{0}^{\textrm{A}}$.}
\label{TabDipFields}%
\bgroup
\def\arraystretch{1.2}
    \begin{tabular}{| c | c | c | c | c | c | c |}
        \hline
        \textbf{Site} & \textbf{Wyckoff position} &
        $x$ \textbf{(a)}, $y$ \textbf{(a)}, $z$ \textbf{(c)}&
        $B_{\mathrm dip}^{\bm{\mu}_{\textrm{Co}} \parallel c}$ (G) &
        $B_{\mathrm dip}^{\bm{\mu}_{\textrm{Co}} \perp c} $ (G) &
        $V_{0}^{i}$ \textbf{(eV)} & $E_{\textrm{ZP}}^{i}$ \textbf{(eV)}\\
        \hline
        \multicolumn{7}{|c|}{{Ambient $P$}
        ($\Delta_{\textrm{AB}} = \Delta_{\textrm{BC}} = 1.42$ eV,
        $\Delta_{\textrm{AC}} = 1.06$ eV)}\\
        \hline
        $A$ & $2c$ &\nicefrac{1}{4} \nicefrac{1}{4} 0.875 & 330 &
        170 & 0 & 0.45\\
        $B$ & $2c$ &\nicefrac{1}{4} \nicefrac{1}{4} 0.42 & 1100 &
        550 & 0.27 & 0.63\\
        $C$ & $4f$ &\nicefrac{3}{4} \nicefrac{1}{4} 0.30 & - & - &
        0.56 & 1.13\\ 
        \hline
        \multicolumn{7}{|c|}{{${P = 30}$ kbar}
        ($\Delta_{\textrm{AB}} = \Delta_{\textrm{BC}} = 1.36$ eV,
        $\Delta_{\textrm{AC}} = 1.08$ eV)}\\
        \hline
        $A$ & $2c$ &\nicefrac{1}{4} \nicefrac{1}{4} 0.875 & 330 &
        170 & 0 & 0.45\\
        $B$ & $2c$ &\nicefrac{1}{4} \nicefrac{1}{4} 0.42 & 1100 &
        550 & 0.37 &  0.73\\
        $C$ & $4f$ &\nicefrac{3}{4} \nicefrac{1}{4} 0.30 & - & - &
        0.62 & 1.22\\
        \hline
    \end{tabular}
\egroup
\end{table*}
Energy-volume curves were obtained from the PW-based calculations by
constant volume energy minimization. The convergence criteria for
forces minimization was set to $5 \times 10^{-3}$ eV/Å and $10^{-5}$
eV was used as threshold for self-consistent electronic cycles. The
optimized unit cell volume at ambient $P$ for LaCoPO is $133.21$
Å$^{3}$ with lattice parameters $a = 3.966$ Å and $c = 8.468$ Å. In
accordance with previous findings,\cite{Yan08} the
density-functional theory (DFT) calculations reproduce the
experimental structural parameters with errors $\sim 1$\%. No
anomalies in the energy-volume curves are observed within the
explored $P$ range. At the same time, the FM-ordered configuration
is the GS for $P$ values up to $100$ kbar in agreement with
experimental results.\cite{Yan08} The results then suggest that both
the crystal structure and the FM configuration as GS are stable
against the increase of $P$ in LaCoPO at least in the investigated
$P$ range. The calculated value for the magnetic moment
$\mu_{\textrm{Co}} = 0.57 \; \mu_{\textrm{B}}$ is in agreement with
previous calculations.\cite{Yan08} This value is slightly higher
than our experimental estimate $\mu_{\textrm{Co}} \simeq 0.3 \;
\mu_{\textrm{B}}$ and it is substantially unchanged for $P < 40$
kbar while, for higher pressures, the magnetic moment on Co atoms
linearly decreases reaching $0.55 \; \mu_{\textrm{B}}$ for $P = 100$
kbar.

\subsection{The interstitial $\mu^{+}$ sites and their stability
against zero-point motion and external pressure}

\begin{figure*}[htbp]
\begin{center}
\includegraphics[scale=0.4]{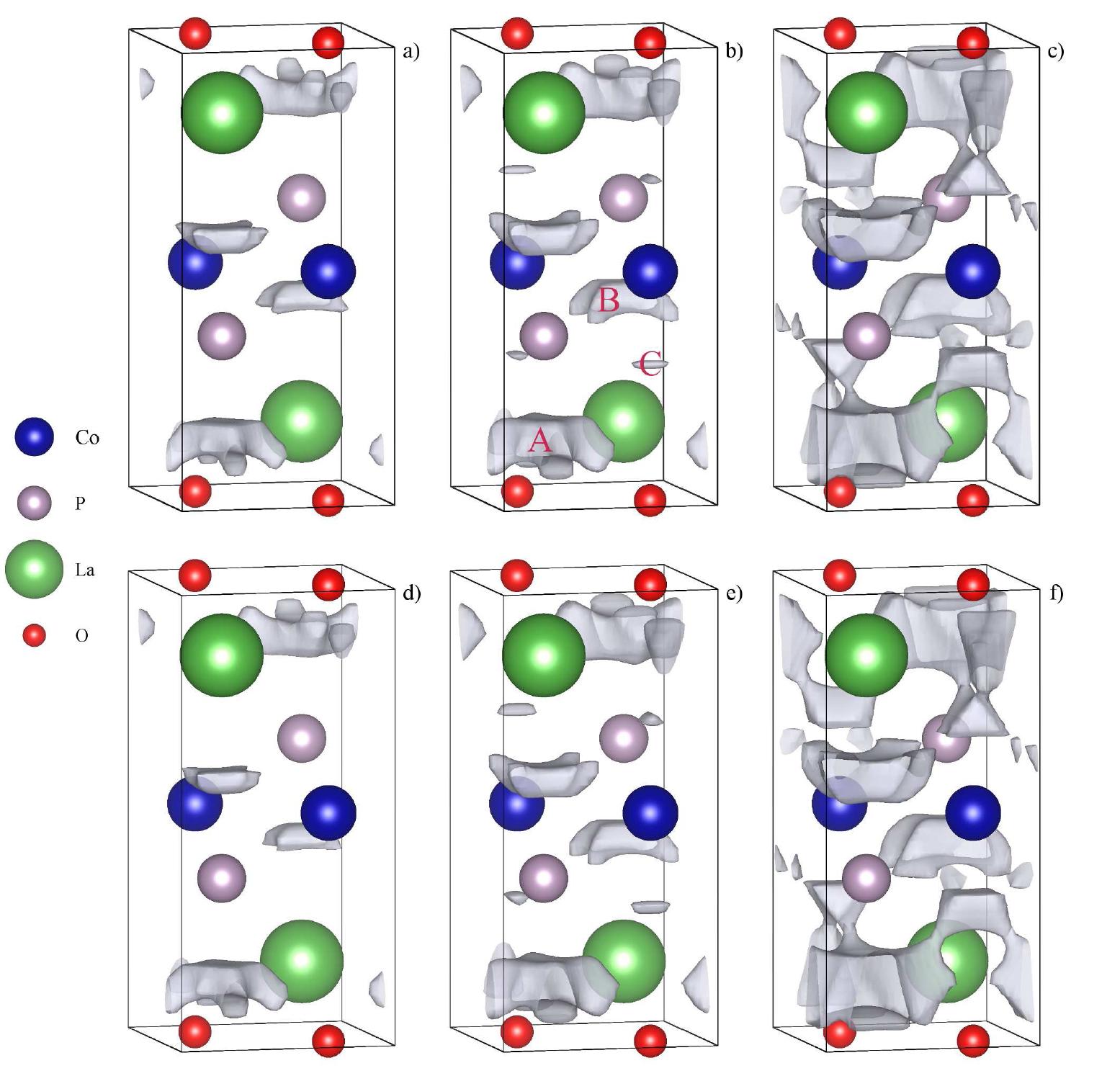}%
\caption{\label{GraMuonPos}(Color online) Isosurfaces of the
electrostatic potential of LaCoPO for ambient $P$ (upper) and $P =
30$ kbar (lower). For each pressure, the energies corresponding to
the first eigenvalue for a muon in site $A$, $B$, and $C$ are shown,
i.e., $V(\mathbf{r})^{\textrm{iso}} = E_{\textrm{ZP}}^{\textrm{A}}$
for a) and d); $V(\mathbf{r})^{\textrm{iso}} =
E_{\textrm{ZP}}^{\textrm{B}}$ for b) and e);
$V(\mathbf{r})^{\textrm{iso}} = E_{\textrm{ZP}}^{\textrm{C}}$ for c)
and f).}
\end{center}
\end{figure*}

The crystallographic sites where $\mu^{+}$ stop after thermalization
processes may be identified to a first extent by calculating the GS
electron density of the bulk material under investigation. The
minima of the electrostatic potential consequently obtained are
indeed found to provide a correct estimate to this aim.\cite{DeR12}
Three inequivalent positions are computed for LaCoPO, whose absolute
values of the potential energy are denoted as $V_{0}^{i}$ ($i =$ A,
B, C) where we conventionally set $V_{0}^{\textrm{A}} \equiv 0$ eV.
Site A (B) is located within the LaO (CoP) tri-layers while site C
is aligned with O and Co in between the different tri-layers, as
shown in Fig. \ref{GraMuonPos}b. Remarkably, we note that the
electrostatic interaction favours site A unlike what is found in
RECoAsO and REFeAsO where the interstitial site close to the
transition metal plane is favored.\cite{DeR12,Sug11} More detailed
information about the crystallographic positions of the three sites
and their evolution upon increasing $P$ is reported in Tab.
\ref{TabDipFields} and in Fig. \ref{GraMuonPos}. The local magnetic
fields at the $\mu^{+}$ sites $\mathbf{B}_{\textrm{dip}}$ arising
from the dipolar contribution of $\mu_{\textrm{Co}} = 0.3 \;
\mu_{\textrm{B}}$ magnetic moments were also computed and reported
in Tab. \ref{TabDipFields} (assuming an undistorted lattice).

Once thermalized, in view of the quite high zero-point (ZP) motion
energy for $\mu^{+}$, the potential minima are not necessarily
stable trapping sites. Whether the ZP energy is higher than the
potential barrier surrounding a minimum, $\mu^{+}$ can escape to
lower-energy ones. To check if the minima yield stable trapping, we
computed the ZP motion by solving the Schr\"odinger equation for the
$\mu^{+}$ in the electrostatic potential. Then, we inspected the
position of the probability maxima for the relative GS wavefunctions
and characterized each $\mu^{+}$ site by the relative lowest
eigenvalue $E_{\textrm{ZP}}$ (see Tab. \ref{TabDipFields}). It
should be remarked that a stable trapping site must have a nodeless
and well-localized GS. Calculations show that this is clearly the
case for sites A and B, while for site C we found a delocalized
wavefunction with a large amplitude around C (resonant state).
Moreover for site C, at variance with the findings for sites A and
B, the lowest eigenvalue $E_{\textrm{ZP}}^{\textrm{C}}$ is higher
than the energy barrier separating A and C. Under these conditions,
all $\mu^{+}$ stopping in C reach the energetically more favorable
minimum A. We cannot exclude the presence of a self trapping
mechanism which could, in principle, prevent $\mu^{+}$ from reaching
the lowest energy site A from minimum C. Anyway, the C site has a so
high energy that the fraction of $\mu^{+}$ stopping there should be
negligible. From the above considerations, only A and B minima are
eligible as stable $\mu^{+}$ sites and C will be disregarded in the
rest of the discussion.

As already mentioned, DFT calculations does not evidence structural
or magnetic transitions triggered by $P$. The energy minima and ZP
energies of the interstitial sites are slightly modified with
increasing $P$ up to $30$ kbar but variations are smooth and of the
order of the estimated accuracy of the calculation. The same trend
shown in Tab. \ref{TabDipFields} for the energy minima and the zero
point motion is found also for higher pressures ($P$ up to 100
kbar). Minima B and C, in particular, increase their energy with
respect to site A while the ZP energies of both the sites (with
reference to the relative energy minimum) do not change
significantly. This allows to state that the occupation probability
of those sites should remain practically unchanged upon increasing
$P$ or, at least, that site A turns out to be even more and more
favourable from an energetic point of view in comparison with B and
C. In conclusion, no evidence for sudden changes in the population
of interstitial sites can be derived from DFT calculations.

After considering both the experimental and the computational
results, the possibility of a sudden change of site occupation for
$\mu^{+}$ upon increasing $P$ seems to be very unlikely, as
described in Sect. \ref{SectFPCalc}. From the experimental side, in
fact, no sign of occupancy of different sites is present for any of
the investigated samples (see the inset of Fig.
\ref{GraInternalFields} for representative raw data). An
unrealistically complete redistribution of the site occupation
between the two inequivalent sites A and B would then be needed in
order to explain the experimental results. As already stressed
above, moreover, no sign of structural transitions upon increasing
$P$ for the investigated samples was evidenced by means of DFT
computations. Due to the smoothness of the energy landscape coming
out from calculations, then, it seems to be highly unlikely to
attribute the sharp and dramatic drop in the internal field
occurring at $P \simeq 5$ kbar (see the main panel of Fig.
\ref{GraIntFieldAndMagMom}) to a modification in $\mathbf{r}_{\mu}$.

\subsection{Computation of the perturbation effect induced by the
muon}

In this paragraph we will consider the local perturbation induced by
the implanted muons. We studied the effect of $\mu^{+}$ induced
lattice distortion on the electrostatic potential landscape.
Normally, the dielectric screening in metals is so efficient that
$\mu^{+}$ will not cause significant lattice distortion. However,
here we are dealing with a material that is a poor metal, and so it
is important to get an estimate of any lattice distortion effect.
Since we need to get just an estimate we can ignore the effect of ZP
motion and treat the $\mu^{+}$ as if it were a hydrogen interstitial
impurity. We model the isolated impurity within the supercell
approach by building a $64$ atoms supercell up from our bulk
structure. The supercell volume was kept constant while atomic
positions were allowed to relax. $\mu^{+}$ are positively charged
particles but, at the same time, our supercell must have a neutral
charge because the electronic screening in metals does neutralize
the $\mu^{+}$ charge. For the hydrogen impurity we chose those sites
identified as absolute minima by the electrostatic potential
landscape. We then let the system evolve towards the GS allowing
both electron rearrangement and lattice distortion. The final
optimized position for the impurity represents the refined site for
the $\mu^{+}$, and we found that the refined $\mu^{+}$ position is
in agreement with the one obtained by analyzing the electrostatic
potential minima. Fully relaxed structures at ambient $P$, at $P =
15$ kbar and at $P = 30$ kbar show that both the $\mu^{+}$ position
inside the cell and the distance between $\mu^{+}$ and P ions varies
less than $0.04$ Å. The phosphorous ion close to $\mu^{+}$ is pushed
closer to the Co plane by $\sim 0.06$ Å and its four neighbouring Co
atoms increase their magnetic moment to $0.6 \; \mu_{\textrm{B}}$.
Nevertheless, once more, no appreciable modification of the crystal
structure and of the magnetic properties of the whole system
(crystal and $\mu^{+}$) which could in principle justify the drop of
the internal field observed around $P \sim 5$ kbar could be computed
as a function of $P$.

\subsection{Pressure-induced effects on the energy bands}

\begin{figure}[t!]
\begin{center}
\includegraphics[scale=0.32]{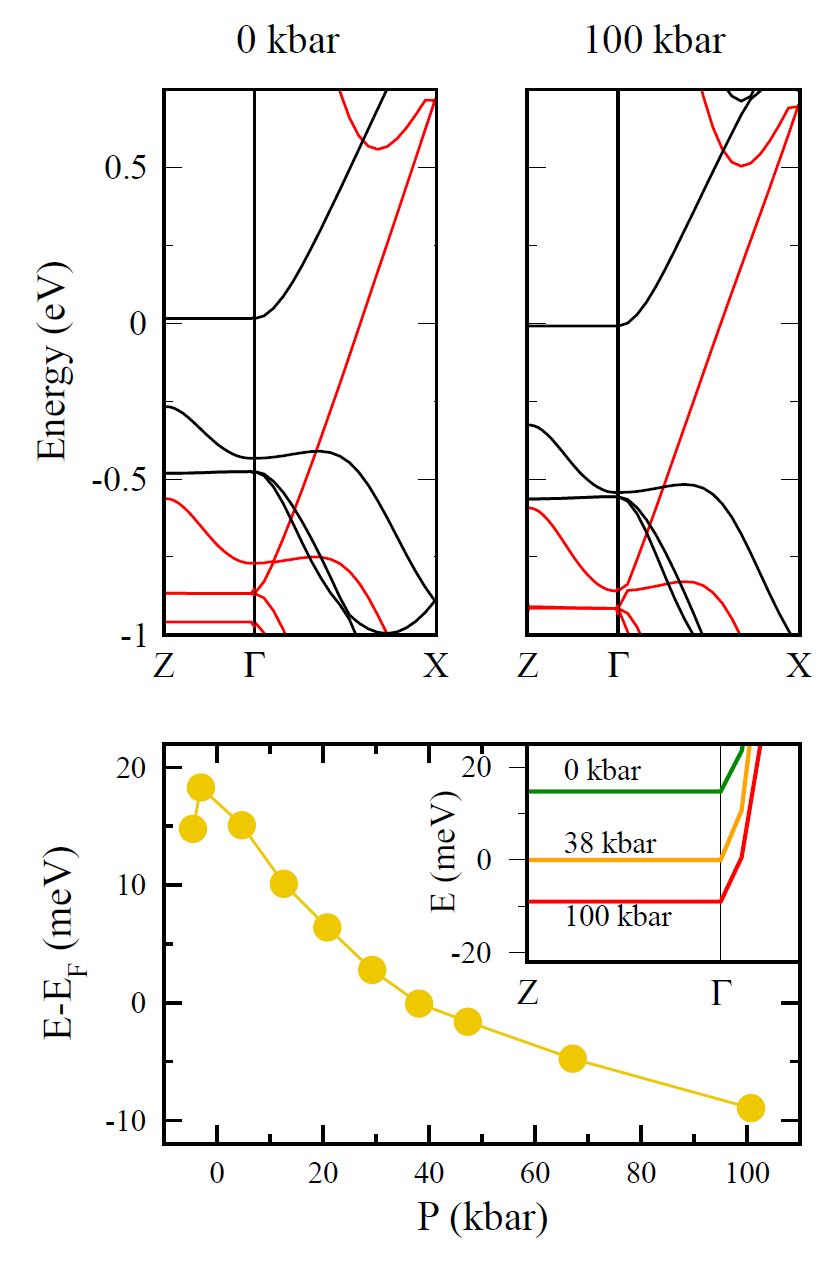}%
\caption{\label{GraBandCrossing} (Color online) Upper panels: energy
bands of LaCoPO at ambient $P$ and at 100 kbar ($E_{F}=0$).
Black(red) color refer to majority (minority) spin bands. Lower
panel: difference between the energy at $\Gamma$ and the Fermi
energy $E_{F}$ for the band crossing the Fermi energy. Inset: energy
band dispersion for selected $P$ values.}
\end{center}
\end{figure}
In order to identify possible changes of the spin density at the
$\mu^{+}$ site we computed the band structures to evaluate the
contribution of the conduction electrons to the hyperfine field as a
function of pressure. At $P \sim 38$ kbar an unoccupied electron
band shifts across the Fermi energy as shown in Fig.
\ref{GraBandCrossing} creating a large cylindrical Fermi surface
owing to the flat dispersionless band along $\Gamma$-Z shown in
figure. This modification in the band structure strongly suggests
that the variation of the local field upon increasing $P$ can be due
to a change in the transferred hyperfine contact field
$\textbf{B}_{\textrm{c}}(\mathbf{r}_{\mu})$ reported in Eq.
\eqref{EqSumLocalFields}, strongly influenced by the conduction
electrons.\cite{Sli90} It should be remarked that, according to what
is reported in Tab. \ref{TabDipFields}, for the in-plane orientation
of the spins one has
$\textbf{B}_{dip}\left(\mathbf{r}_{\mu}(\textrm{A})\right) = -
\textbf{B}_{L}$ and the muon in site A is thus subject only to the
contribution of $\textbf{B}_{\textrm{c}}(\mathbf{r}_{\mu})$.

\section{Discussion}

Both chemical and external pressure have been shown to affect the
ferromagnetic properties of the Co-based $1111$ oxy-pnictides.
ZF-$\mu^{+}$SR experiments display that the magnetic ordering
temperature $T_{\textrm{c}}$ increases as a function of pressure.
The internal field at the muon site $B_{\mu}$ is found to abruptly
decrease around $5$ kbar in LaCoPO. A change of the same amount is
found when lanthanum is substituted by smaller Pr ions. This
demonstrates that chemical pressure mimics the hydrostatic pressure
as can be deduced from Fig. \ref{GraIntFieldAndMagMom}. Claims of
close analogies among the effect of chemical and external pressures
were already reported concerning both $1111$ and $122$
compounds.\cite{Kim09,Pra12b,Gat12} In order to understand the
origin of the drastic reduction observed for $B_{\mu}$ we have
investigated the behavior of those quantities, implicitly included
in Eq. \eqref{EqSumLocalFields}, that mainly contribute to the
precession frequency of the muon around the local field, namely the
Co ordered magnetic moment, $\mu_{\textrm{Co}}$, and both the
distance and the density of the spin at the $\mu^{+}$ site as a
function of $P$.

Magnetization measurements have shown no significative change of
$\mu_{Co}$ neither by changing chemical or external pressure (see
the inset of  Fig. \ref{GraSQUID}). DFT calculations have been
performed in order to determine the muon sites and displays that
only two lattice positions are stable, namely close to the CoP (site
A) and to the REO (site B) layers (Fig. \ref{GraMuonPos}), the
latter being energetically favored. Since ZF-$\mu^{+}$SR display
only one oscillating component, only one site is really occupied.
The calculation of electrostatic potentials does not display any
strong variation as a function of pressure, indicating that the
$\mu^{+}$ site cannot change. Since also $\mu_{\textrm{Co}}$ is
constant, we can conclude that neither the dipolar interaction or
the Lorenz field of Eq. \eqref{EqSumLocalFields} can be responsible
for the suppression of the local field at the $\mu^{+}$ site as a
function of pressure. On the other hand, DFT band structure
calculations show that a slight variation in the energy of a
minority band occurs as a function of pressure, displaying that a
crossing of the Fermi level takes place around $38$ kbar, as shown
in Fig. \ref{GraBandCrossing}. This computational outcome suggests a
sudden change of the hyperfine contribution in Eq.
\eqref{EqSumLocalFields} due to the occupation of this minority spin
band. A notable discrepancy between the critical $P$ value for the
drop of the internal field from experimental ($\sim 5$ kbar) and DFT
($\sim 38$ kbar) results is found. This difference may be understood
by considering that, first, the band structure is extremely
sensitive to the position of the P ions. The shift of the position
of P ions towards the Co plane introduced by the incoming $\mu^{+}$
discussed above is not taken into account when inspecting the band
structure and it may indeed favour the population of the unoccupied
valence band at lower $P$ values. In addition, in this delicate
scenario, even small non-hydrostatic effects not taken into account
in the calculations can shift the critical pressure. Moreover, the
DFT approach for the first-principles analysis may underestimate $P$
effects as a consequence of the overestimated magnetic moment which
in turn is related to the height of phosphorous ions.\cite{Zha09}
Unfortunately, as already recalled above, no experimental structural
refinements as a function of $P$ are available at the moment. Thus,
we cannot quantify the mismatch between DFT obtained and real
crystal structures as a function of pressure.

Other hypotheses can be made in order to explain the observed
phenomenology by referring to the intrinsic magnetic properties of
the materials, whose changes upon increasing $P$ can be directly
reflected in relative modifications into one (or more) terms
appearing in Eq. \eqref{EqSumLocalFields}. One can then take into
consideration the possibility of a $P$-triggered reorienting of the
FM configuration. The values for the modification of the overall
local magnetic field at the muon site according to this hypothesis
are provided in Tab. \ref{TabDipFields}. Besides a possibly
satisfying explanation of experimental data from a quantitative
point of view, this hypothesis should be considered as unrealistic
mainly due to the low absolute value of spin-orbit interaction for
the Co ions. The different energies of the GSs were analyzed by
performing non-collinear spin-polarized calculations for different
$P$ and easy magnetization axes orientations. The results for
different spin configurations are found to be nearly degenerate so
that, in conclusion, the mean-field approach does not provide a clue
to support this alternative.

\section{Conclusions}

We performed detailed measurements by means of muon spin
spectroscopy on LaCoPO, PrCoPO and LaCoAsO under applied hydrostatic
pressure. The localized electrons on the ionic shells of Pr$^{3+}$
ions do not affect at all the local static features of magnetism as
detected by muons in spite of the high value of the magnetic moment
expected for the free Pr$^{3+}$ ion. Phosphorous-based compounds
turn out to be much more sensitive than LaCoAsO to the application
of pressure. In particular, the critical transition temperature to
the ferromagnetic phase in LaCoPO is sizeably increased both by
chemical and external pressures (the former being triggered by the
full Pr$^{3+}$/La$^{3+}$ substitution), while for LaCoAsO the effect
is also present but by far less marked. The increase of both kinds
of pressure, moreover, dramatically suppress the local magnetic
field at the muon site leaving the magnetic moment per Co ion
substantially unchanged. Density-functional theory calculations of
the band structure suggest that this the change of the muon field is
due to a subtle change of the hyperfine coupling driven by a
minority band crossing the Fermi surface. Results clearly evidence
how crucial the role of pressure is on the properties of the
ferromagnetic phase and, at the same time, how similar the effects
of chemical and external hydrostatic pressure is in $1111$
oxy-pnictide compounds. A computational description of LaCoPO
performed by means of \emph{ab-initio} DFT calculations supports
these findings and suggests that chemical and external pressures
both trigger a qualitative change in the electronic band structure
of the investigated compounds.

\section{Acknowledgements}

G. Prando acknowledges support from the Leibniz-Deutscher
Akademischer Austauschdienst (DAAD) Post-Doc Fellowship Program. G.
Profeta acknowledges support from the FP7 European project
SUPER-IRON (grant agreement No. 283204), by a CINECA-HPC ISCRA grant
and by an HPC grant at CASPUR. F. B. acknowledges support from
CASPUR under the Standard HPC Grant 2012 and from the FP7 European
project SUPER-IRON (grant agreement No. 283204). E. M. B. and H.-J.
G. acknowledge support from Deutsche Forschungsgemeinschaft (DFG)
through SPP1458 (grant No. GR3330/2). P. C. and S. S. acknowledge
support from Fondazione Cariplo (research grant No. 2011-0266). The
structural models in the figures were rendered using the VESTA
package.\cite{Mom11}



\end{document}